\documentclass[%twocolumn,
secnumarabic, amssymb, amsmath, nofootinbib,tightenlines,showpacs,
nobibnotes, aps, prl]{revtex4}
\usepackage{amsfonts}
\usepackage{docs}%
\usepackage{bm}%
\usepackage[colorlinks=true,linkcolor=blue]{hyperref}%
\usepackage{graphicx}% Include figure files
\usepackage{dcolumn}% Align table columns on decimal point

\begin{document}

\title{Topological Phase Separation In Trapped Ultracold Fermionic Gases}

\author{Xiaosen Yang}
\author{Shaolong Wan}
\email[]{slwan@ustc.edu.cn} \affiliation{Institute for Theoretical
Physics and Department of Modern Physics University of Science and
Technology of China, Hefei, 230026, \textbf{P. R. China}}
\date{\today}

\begin{abstract}
We investigate the harmonically trapped 2D fermionic systems with
a effective spin-orbit coupling and intrinsic s-wave
superfluidity under the local density approximation, and find
that there is a critical value for Zeeman field. When the Zeeman
field larger than the critical value, the topological
superfluid phases emerge and coexist with the normal
superfluid phase, topological phase separation, in the
trapped region. Otherwise, the superfluid phase is
topologically trivial.
\end{abstract}


\maketitle

\section{Introduction}

Recently, topological superconductors/superfluids(TSCs/TSs) which cannot be
described by the Landau symmetry breaking theory, but the
topological order \cite{X. G. Wen}, have attracted a great
attentions in condensed matter physics \cite{N. Read, X.l.Qi,
Zhong Wang1, M sato1, schnyder, Kubasiak, Tewari}. TSCs which
resemble topological insulators(TI) \cite{Bernevig, Hasan, Zhong
Wang2, C. L. Kane}, have gapped bulk and gapless Majorana edge
states on the boundaries which are protected by the topological
properties of the bulk for the bulk-edge correspondence. The
distinguishable phases have same symmetry but cannot change into
each other without topological phase transition(TPT). The
realization of TSCs/TSs has been proposed in a large spin-orbit
coupling(SOC) semiconductors with a proximity induced s-wave
pairing potential and ultracold atomic gases with intrinsic
superfluidity in the presence of a Zeeman field and an effective
SOC.

Ultracold atomic gases with remarkable controllability provide an
ideal stage to investigate the phenomena in the condensed matter
physics \cite{Schunck, N. Gemelke, Simons, G.Lamporesi, Recati,
Giorgini, Bloch, Bedaque}. The effective SOC which is essential
for the realization of different TIs and TSCs/TSs, can be generated by
utilizing the spatial varying laser fields \cite{S. L. Zhu, N.
Goldman, K. Osterloh} and has been realized in experiment \cite{Y.
J. Lin1, Y. J. Lin2}. The topological properties of the
superconductors/superfluids were investigated in \cite{Stanescu, M
sato1, M sato2}. The SOC can change the ground state of the boson
condensate \cite{Stanescu1, C. Wang}, too.

In this paper, we investigate the topological properties of 2D 
s-wave superfluids trapped in a harmonic potential with a SOC
in local-density approximation(LDA). The superfluid phase is
topologically trivial in the absence of the Zeeman field. An
effective Zeeman field, breaking the time-reverse symmetry and
induced by the spin imbalance of the population, is needed to
realize the topologically nontrivial phases. The topological
nontrivial phases emerge in the trapped region when the Zeeman
field is larger than the critical value. There is topological
phase transition when the Zeeman field crosses the critical
values. In the presence of the harmonic potential, the chemical
potential and the pairing gap are functions of the position
coordinate. Such that the topological properties of the
superfluids are also changing with the coordinate. Analogy to
the case of without the SOC \cite{Simons, Bedaque, Sheehy, Perali,
Gubbels}, we find there exist topological phase separation(TPS)
phenomena which is the coexistence of topological trivial and
nontrivial superfluid phases in the trapped region. The
gapless Majorana edge states are localized at the interfaces
between the topologically superfluid phases (TS) and the normally superfluid phases (NS) in the
trapped region.

This paper is organized as follows. In Sec. II, first, we
introduce the Hamiltonian of the model. Then, we derive the energy
gap closing conditions and introduce the TKNN numbers. In Sec.
III, we investigate the topological properties of the
superfluid phase and find there exist topological phase
separation phenomena in the trapped region. We also extend the
analysis to the continuum model. The conclusions are given in Sec.
IV.

\section{Formalism of the System}

We consider spin-singlet superfluids with the Rashba SOC on a
2D square optical lattice in the harmonic potential, which is
described by the Hamiltonian($\hbar=k_{B}=1$):

\begin{eqnarray}
H=H_{0}+H_{SO}+H_{s},
\end{eqnarray}
where $H_{0}$ is the kinetic term with a Zeeman field, $H_{SO}$ is
the spin-orbit interaction term, and $H_{s}$ is the pairing
potential term. They take the following forms
\begin{eqnarray}
H_{0}&=&-t \sum_{{\bf i},{\bf r},\sigma}\sum_{\hat{e}=\hat{x},\hat{y}} (c_{{\bf i}+\hat{e}\sigma}^{\dag}c_{{\bf i}\sigma}+h.c.) -\sum_{{\bf i},{\bf r}}[(\mu(r)+h)c_{{\bf i}\uparrow}^{\dag}c_{i\uparrow} +(\mu(r)-h)c_{i\downarrow}^{\dag}c_{i\downarrow}],\\
H_{SO}&=&-\lambda\sum_{{\bf i},{\bf r}}[(c_{{\bf i}-\hat{x}\downarrow}^{\dag}c_{{\bf i}\uparrow}
- c_{{\bf i}+\hat{x}\downarrow}^{\dag}c_{{\bf i}\uparrow})
+ i(c_{{\bf i}-\hat{y}\downarrow}^{\dag}c_{{\bf i}\uparrow}
- c_{{\bf i}+\hat{y}\downarrow}^{\dag}c_{{\bf i}\uparrow}) + h.c.],\\
H_{s}&=&-\sum_{{\bf i},{\bf r}}\Delta(r)(c_{{\bf i}\uparrow}^{\dag}c_{{\bf i}\downarrow}^{\dag}+h.c.),
\end{eqnarray}
where $c_{{\bf i},\sigma}^{\dag}(c_{\bf{i},\sigma})$ denotes the
creation (annihilation) operator of the fermion with spin
$\sigma=(\uparrow,\downarrow)$ at site ${\bf i}=(i_{x},i_{y})$,
$\lambda$ is the strength of SOC and $\Delta(r)$ is the pairing
potential of the superfluid. $\mu(r)=\mu-V(r)$ is the local
chemical potential, here, $\mu$ is global chemical potential and
$V(r)=m\omega^{2} r^{2}/2$ is harmonic potential. $h$ is a Zeeman
field. Introduce $\Psi_{{\bf k}, r}^{\dag}=(c_{{\bf
k},\uparrow}^{\dag}, c_{{\bf k},\downarrow}^{\dag},c_{-{\bf
k},\uparrow},c_{-{\bf k},\downarrow})$ with $c_{{\bf
k}}^{\dag}=\sum_{{\bf i}}e^{i{\bf k}\cdot{\bf i}}c_{{\bf
i}}^{\dag}$, the Hamiltonian, in momentum space, can be rewritten
as
\begin{eqnarray}
H=\frac{1}{2}\sum_{{\bf k},{\bf r}}\Psi_{{\bf k},r}^{\dag}\mathcal{H}({\bf k},r)\Psi_{{\bf k},r},
\end{eqnarray}
with
\begin{eqnarray}
\mathcal{H}({\bf k},r)=\left(
  \begin{array}{cc}
    \epsilon({\bf k},r)-h \sigma_{z} + {\bf g}_{{\bf k}} \cdot {\bf \sigma} & i \Delta({\bf k},r) \sigma_{y}\\
    - i \Delta({\bf k},r)^{\ast} \sigma_{y} &  -\epsilon({\bf k},r) + h \sigma_{z} + {\bf g}_{{\bf k}} \cdot {\bf \sigma}^{\ast}\\
  \end{array}
\right),
\end{eqnarray}
where ${\bf \sigma}=(\sigma_{x},\sigma_{y})$ are Pauli matrices,
$\epsilon({\bf k},r)= -2t(\cos k_{x} + \cos k_{y})-\mu(r)$, and
${\bf g}_{{\bf k}}=2\lambda(\sin k_{y},  -\sin k_{x})$.
Diagonalizing the Hamiltonian, we find the eigenvalues $E({\bf
k},r)$ of the Hamiltonian satisfy:
\begin{eqnarray}
E({\bf k},r)^{2}=\epsilon({\bf k},r)^{2}+h^{2}+|{\bf g}_{{\bf
k}}|^{2} +\mid\Delta({\bf k},r)\mid^{2} \pm 2
\sqrt{h^{2}(\epsilon({\bf k},r)^{2}+ \mid\Delta({\bf
k},r)\mid^{2}) +\epsilon({\bf k},r)^{2} |{\bf g}_{{\bf k}}|^{2}}.
\end{eqnarray}

The pairing potential and the global chemical potential can be
tuned by the inter-species interaction via the Feshbach resonance
and the filling number. The pairing potential is a smooth function
of the position coordinate. We consider the Zeeman field $h\geq0$.
In the presence of pairing ($\Delta({\bf k},r)\neq0$), the
spectrum of our model is gapless only when $|{\bf g}_{{\bf k}}|=0$
at four momenta ${\bf k}_{0}=(0,0),(0,\pi),(\pi,0),(\pi,\pi)$, and
the Zeeman magnetic field must satisfies $h=\sqrt{\epsilon({\bf
k}_{0},r)^{2}+|\Delta({\bf k_{0}},r)|^{2}}$ which can be written
as:
\begin{eqnarray}
     \left\{
          \begin{array}{c}
            h_{1}=\sqrt{(4t+\mu(r))^{2}+|\Delta({\bf k}_{0},r)|^{2}}, ~~~~ {\bf k_{0}}=(0,0)\\
            h_{2}=\sqrt{\mu(r)^{2}+|\Delta({\bf k}_{0},r)|^{2}},~~~~~~~~~ {\bf k_{0}}=P[(\pi,0)]\\
            h_{3}=\sqrt{(4t-\mu(r))^{2}+|\Delta({\bf k}_{0},r)|^{2}}, ~~~~{\bf k_{0}}=(\pi,\pi)\\
          \end{array}
     \right.,
\end{eqnarray}
where $P[{\bf k}]$ is a set of the vectors obtained from
permutations of vector ${\bf k}$.

We know that this BdG Hamiltonian has particle-hole symmetry(PHS)
\begin{eqnarray}
\mathcal{H}({\bf k},r)=- t_{x}\mathcal{H}(-{\bf k},r)^{T}t_{x},
\end{eqnarray}
where $t_{x}$ is Pauli matrix. The classification of time-reversal
breaking(TRB) BdG Hamiltonian with PHS is $\mathbb{Z}$ class
\cite{schnyder}. The topological numbers which characterize the
topological properties of the superfluid phases are integers.
The Hamiltonian has four energy bands, only two of them can be
gapless and others are always gapped. Only the gapless energy
bands have contribution to the topological number. We consider the
s-wave pairing case with a real pairing gap function $\Delta({\bf
k},r)=\Delta_{s}(r)$. then, we can deform the Hamiltonian
$\mathcal{H}$ into a dual Hamiltonian $\mathcal{H}^{D}$ by a
unitary transformation
\begin{eqnarray}
\mathcal{H}^{D}({\bf k},r)=D \mathcal{H}({\bf k},r) D^{\dagger},
\end{eqnarray}
where, the dual Hamiltonian and the unitary matrix $D$ is given
as:
\begin{eqnarray}
\mathcal{H}^{D}({\bf k},r)=\left(
  \begin{array}{cc}
    \Delta({\bf k},r)-h\sigma_{z} &
    -i\epsilon({\bf k},r)\sigma_{y}-i{\bf g}_{{\bf k}} \cdot {\bf \sigma} \sigma_{y} \\
    i\epsilon({\bf k},r)\sigma_{y} +i{\bf g}_{{\bf k}} \sigma_{y} \cdot {\bf \sigma} &  \Delta({\bf k},r)+h\sigma_{z}\\
  \end{array}
\right),\label{dualh}
\end{eqnarray}
and
\begin{eqnarray}
D= \frac{1}{\sqrt{2}}\left(
  \begin{array}{cc}
    1 & i \sigma_{y} \\
    i \sigma_{y} &  1\\
  \end{array}
\right).
\end{eqnarray}

We can adiabatically deform the Hamiltonian by taking
$\epsilon({\bf k},r)=0$ without close the energy gap. Such that
the dual Hamiltonian is decomposed into the following two
$2\times2$ parts for dual up-spin and down-spin,
\begin{eqnarray}
\mathcal{H}_{\uparrow \uparrow}^{D}({\bf k},r) = \left(
  \begin{array}{cc}
    \Delta_{s}(r)- h & 2 \lambda ( \sin k_{y} + i \sin k_{x}) \\
    2 \lambda ( \sin k_{y} - i \sin k_{x}) &  - \Delta_{s}(r)+h \\
  \end{array}
\right),
\end{eqnarray}
\begin{eqnarray}
\mathcal{H}_{\downarrow \downarrow}^{D}({\bf k},r) = \left(
  \begin{array}{cc}
    \Delta_{s}(r)+h & 2 \lambda ( -\sin k_{y} + i \sin k_{x}) \\
    2 \lambda ( -\sin k_{y} - i \sin k_{x}) &  -\Delta_{s}(r)-h \\
  \end{array}
\right).
\end{eqnarray}

When the Zeeman field $h>0$, the energy gap of the dual up-spin
can vanish while the down-spin is always gapped. And when the
Zeeman field $h<0$, the reverse case happen. For the
simplification, we only consider $h>0$ case and rewrite the
Hamiltonian of the dual up-spin as:
\begin{eqnarray}
\mathcal{H}_{\uparrow\uparrow}^{D}({\bf k},r)={\bf d}({\bf k},r) \cdot {\bf \sigma},
\end{eqnarray}
with ${\bf d}({\bf k})=(2 \lambda \sin k_{y}, -2 \lambda \sin k_{x}, \Delta_{s}(r)-h)$.

As for the harmonic potential, the topological properties of the
superfluid phases change with the distance away from the
trapping center as well as the BdG Hamiltonian. When the harmonic
potential is weak the LDA is reliable. In order to characterize
the topological properties of the phases in the trapped region,
the TKNN number $I_{TKNN}$, The TKNN number $I_{TKNN}$ which is
equivalent to the first Chern number, is introduced,
\begin{eqnarray}
I_{TKNN}= \frac{1}{4 \pi}\int d{\bf k}{\hat {\bf d}}({\bf k})
\cdot \frac{\partial {\hat {\bf d}}({\bf k})}{\partial{k_{x}}}
\times \frac{\partial {\hat {\bf d}}({\bf k})}{\partial{k_{y}}},
\end{eqnarray}
with ${\hat {\bf d}}({\bf k})= {\bf d}({\bf k})/|{\bf d}({\bf
k})|$. For the TRB system, the TKNN number is equals to zero for
the normal phase and nonzero for the topological phase. To
simplify the calculation, we take $\Delta_{s}(r)=\Delta_{s}$.
Considering the Hamiltonian with the fixed pairing gap
$\Delta_{s}$ and a positive tunable Zeeman field $h$, the TKNN
number $I_{TKNN}(h)$ can change only if the Hamiltonian crosses
the gapless point where $d^{2}({\bf k},h)=0$. We find that the
phase is topologically trivial($I_{TKNN}(0)=0$) in the absence of
a Zeeman field. So we only need to calculate the change in
$I_{TKNN}(h)$ at critical points which are $h=\Delta_{s}$ at the
momenta ${\bf k}_{c}=(0,0), P[(\pi,0)], (\pi,\pi)$. We must keep
in mind that the different momenta ${\bf k}_{c}$ indicate the
different adiabatic deformation $\epsilon({\bf k}_{c},r)=0$. Such
that the three critical momenta correspond to three gapless
conditions.

Around the critical momentum ${\bf k}_{c}=(0,0)$, the vector ${\bf
d}({\bf k})$ has an approximate form ${\bf d}({\bf k})=(2 \lambda
k_{y}, -2 \lambda k_{x}, \Delta_{s}(r)-h)$. Taking a cutoff
$\Lambda < \pi$ and in the limit $h- \Delta_{s} = \delta h \ll
\Delta_{s}$, the TKNN number expression can be divided into low
and high energy parts,
\begin{eqnarray}
I_{TKNN}&=& \frac{1}{4 \pi}\int d{\bf k} \frac{{\bf d}({\bf
k})}{d^{3}({\bf k})} \cdot \frac{\partial {\bf d}({\bf
k})}{\partial{k_{x}}} \times \frac{\partial {\bf d}({\bf
k})}{\partial{k_{y}}}\nonumber\\
&=& - \frac{1}{\pi}(\int_{| {\bf k} | \leq \Lambda }d {\bf k} +
\int_{| {\bf k} | > \Lambda} d {\bf k} ) \frac{\lambda^{2} \cos
k_{x} \cos k_{y} \delta h}{(4 \lambda^{2}(\sin^{2} k_{x} +
\sin^{2} k_{y}) + \delta h^{2})^{\frac{3}{2}}}\nonumber\\
&=& I_{TKNN}^{(1)}(\delta h, \Lambda) + I_{TKNN}^{(2)}(\delta h,
\Lambda).
\end{eqnarray}
Since there is no gapless point in the high momentum region $|
{\bf k} | > \Lambda$, the change in $I_{TKNN}$ only come from the
low energy part $I_{TKNN}^{(1)}$. The change in $I_{TKNN}$ can be
calculated at critical momentum ${\bf k}_{c}=(0,0)$,
\begin{eqnarray}
\Delta I_{TKNN}^{(1)}|^{\delta h = 0^{+}}_{\delta h = 0^{-}} = -1,
\end{eqnarray}

Similarly, the change in TKNN number at other critical points can
be obtained and given as:
\begin{eqnarray}
\Delta I_{TKNN}|^{\delta h = 0^{+}}_{\delta h = 0^{-}}= \left\{
     \begin{array}{c}
             -1, ~~~~~~{\bf k}_{c} = (0,0)   \\
             2, ~~~~~~{\bf k}_{c} = P[(0,\pi)] \\
             -1, ~~~~~~{\bf k}_{c} = (\pi,\pi) \\
     \end{array},
     \right.
\end{eqnarray}

\section{Topological Phase Separation}

In the presence of the harmonically trapping potential, the
pairing gap is a function of the distance away from the trapped
center which can be self-consistently determined with the global
chemical potential via the minimization of the free energy. By
using the change in $I_{TKNN}$ at the critical points, we can
study the topological properties of the superfluid phase in
the trapped region. The change in $I_{TKNN}$ at the boundaries can
be listed as below:
\begin{eqnarray}
\Delta I_{TKNN}|^{\delta h = 0^{+}}_{\delta h = 0^{-}}= \left\{
     \begin{array}{c}
     -1,  ~~~~  h=h_{1} \\
     2,  ~~~~~~  h=h_{2}\\
     -1,  ~~~~  h=h_{3}\\
     \end{array}.
     \right.
\end{eqnarray}

In the absence of the Zeeman field, the TKNN number vanish such
that the superfluid phases is topological trivial in the
entire trapped region. As the Zeeman field increasing, the TKNN
number change to nonzero at the critical Zeeman field $h_{c}= {\rm
min}(h_{1}, h_{2},h_{3})$. Such that the topologically nontrivial
phase emerge in the trapped region. There is topological phase
transition when the Zeeman field crosses the critical value. When
the Zeeman field is larger than the critical value, the
topologically trivial and nontrivial phases coexist in the trapped
region. This topological phase separation is analogy to the phase
separation phenomena \cite{Gubbels} in the absence of SOC case.

The interfaces between the superfluid phases of different
topological properties are localized in the trapped region. From
the bulk-edge correspondence, a topologically nontrivial bulk
guarantees the existence of topologically stable gapless edge
states on the interfaces. For this 2D superfluids breaking the
time-reversal symmetry but having the PHS, the gapless edge states
are chiral Majorana fermion modes. The number of gapless edge
states is defined by the TKNN number $I_{TKNN}$. For the odd
$I_{TKNN}$ phase, there are odd Majorana fermions in each vortices
which are neither bosons nor fermions but non-Abelian anyons and
obey the non-Abelian statistics. For the even $I_{TKNN}$ phase, as
for the vortices having even Majorana fermions, there are no
non-Abelian anyons.

Now, we assume the pairing gap is independent of the distance away
from the trapped center ($\Delta_{s}(r)=\Delta_{s}$) and the
harmonic potential is $V(r)=m\omega^{2} r^{2}/2 =0.1 tr^{2}$ where
$r$ is in the unit of the lattice space. We can obtain the phase
diagrams on the $h-r$ plane with the s-wave pairing gap
$\Delta_{s}=t$ for the different global chemical potential, shown
in Fig.[1].

\begin{figure}
\includegraphics[width=9cm]{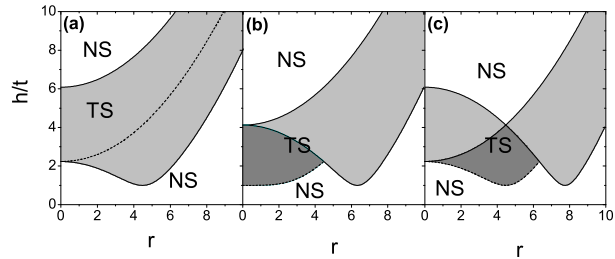}
\caption{We show the phase diagrams on the $h-r$ plane for the
pairing gap $\Delta_{s}=t$, and the global chemical potential (a)
$\mu=-2t$ , (b) $\mu=0$, and (c) $\mu(r)=2t$. The phases of the
shadow regions are TS and others are NS. The TKNN number
$I_{TKNN}$ is $\pm1$ for the slight shadow regions and $2$ for the
dark shadow regions. The entire trapped region is NS phase for
$h/t<1$. As the Zeeman field increasing, there is topological
phase transition at $h/t=1$. The TS phases emerge and coexist
with the NS for the Zeeman field $h/t>1$.}\label{FIG.1}
\end{figure}

\begin{figure}
\includegraphics[width=8cm]{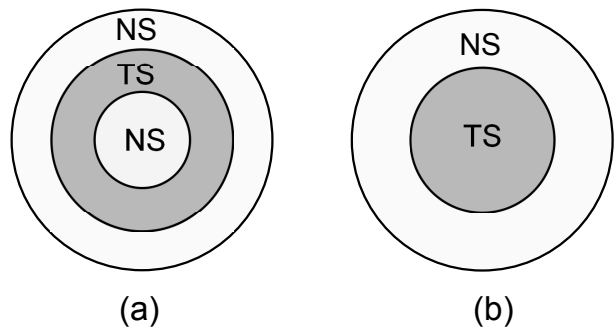}
\caption{We show the diagrams of different types TPS phenomena in real space.
The TPS of (a) is N-T-N and (b) is T-N case in the trapped
region. The phases in the shadow regions are the TS, and the other
regions are the TS.}\label{FIG.2}
\end{figure}

\begin{figure}
\includegraphics[width=9cm]{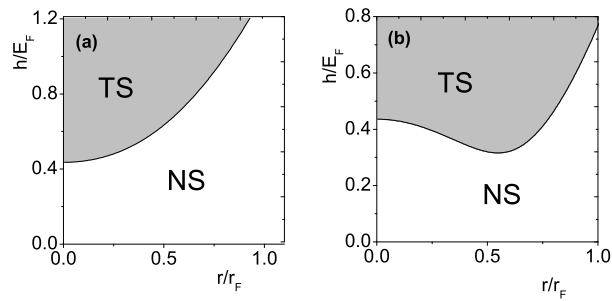}
\caption{We show the phase diagrams on the $h-r$ plane for the
continuum model with the pairing gap $\Delta_{s}=0.1E_{F}$, and
the global chemical potential (a) $\mu=-0.3E_{F}$ and (b)
$\mu=0.3E_{F}$. The phases of the shadow regions are TS and
others are NS. The entire trapped region is NS phase for
$h/t<1$. The TS phase emerges and coexists with the NS for the
Zeeman field $h> min_{r} (\sqrt{\mu(r)^{2} + |\Delta_{s}(r) |^
{2}})$.}\label{FIG.3}
\end{figure}

From Fig.[1], we find that the Zeeman field has a critical value
at $h = t$. When the Zeeman field $h<t$, the superfluid phase
is topological trivial in the entire trapped region. As the Zeeman
field increasing, there is TPT at $h/t=1$. When the Zeeman field
$h>t$, the topologically superfluid phases emerge and
coexist with the normally superfluid phases in the
trapped region. The superfluid phases are topologically
nontrivial for the shadow regions with the others are topologically
trivial. The TKNN number $I_{TKNN}$ is $\pm1$ for the slight
shadow regions and $2$ for the dark shadow regions. When the
global chemical potential $\mu=-2t$, shown in Fig.[1a], there is
TPS phenomena in the trapped region for $h>t$. The TS circle is
immersed in the NS disk for the Zeeman field $1<h/t<2$ and
$h/t>6$. The phase diagram in the real space is N-T-N as shown in
Fig.[2a]. When the Zeeman field $2<h/t<6$, the TS disk is
surrounded by NS circle shown in Fig.[2b]. For $\mu=0$ case,
there is TS disk with $I_{TKNN}=2$ when the Zeeman field
$1<h/t<4$, shown in Fig.[1b], and the TPS phenomena occur only
when the Zeeman field $h>t$. Fig.[1c] gives the phases diagram on
the $h-r$ plane for the global chemical potential $\mu=2t$. The
TS phases emerge and coexist with the NS phase only when $h>t$,
too.

In Fig.[2], we show two types TPS phenomena in the real space.
There are two interfaces between the NS and TS in Fig.[2a] and
one in Fig.[2b]. The interfaces are Majorana gapless edge states.

The TPS is a ubiquitous phenomena in the weak trapped system with
SOC. We extend the above discussion to the continuous model. For
the continuum s-wave case, the Hamiltonian can be written in the
form of Eq.(5) with $\epsilon({\bf k},r)= \hbar^{2} k^{2}/ 2
m-\mu(r)$, ${\bf g}_{{\bf k}}=\alpha(k_{y}, -k_{x})$ and
$\Delta({\bf k},r) = \Delta_{s}(r)$. The gapless conditions are
${\bf k}=0$ and $h= \sqrt{\mu(r)^{2}+ |\Delta_{s}(r)|^{2}}$. There
is TPT at the critical Zeeman field ($h_{c} = min_{r}(
\sqrt{\mu(r)^{2} + | \Delta_{s}(r) |^{2}})$). The TS emerge and
coexist with the NS when $h> h_{c}$. The topological phase
diagrams on $h-r$ plane are shown in Fig.[3a] for $\mu=-0.3E_{F}$
and Fig.[3b] for $\mu=0.3E_{F}$.

\section{Conclusions}

In this paper, we investigate the trapped 2D fermionic systems
with a effective SOC and intrinsic s-wave superfluidity, and
find that the Zeeman field exists a critical value. There is
topological phase transition at the critical Zeeman field. When
the Zeeman field larger than the critical value, the topological
superfluid phase emerge and coexist with the normal
superfluid phase in the trapped region which we called
topological phase separation. Otherwise, the superfluid phase
is topologically trivial. We extend the analysis to the continuum
case and find that the Zeeman field also exists a critical value.
For the continuum case, the topologically superfluid phases
emerge and coexist with the normally superfluid phase When the
Zeeman field larger than the critical value. Otherwise, the
superfluid phase is topologically trivial, too.

\section{Acknowledgments}

This work is supported by NSFC Grant No.10675108.

\end{document}